\documentclass[runningheads]{svmult}

\usepackage{makeidx}   
\usepackage{graphicx}  
\usepackage{subeqnar}  
\usepackage{multicol}  
\usepackage{physprbb}  
\makeindex             

\begin{document}

\title*{~~~~~~~~~~~~~~~~~The TESIS project
~~~~~~~~~~~~~~~~~~~~~~~ Revealing massive early-type galaxies at $z>1$}
\toctitle{The TESIS project:
 Revealing massive early-type galaxies at $z>1$}
%
%
\titlerunning{Revealing massive early-type galaxies at $z>1$}
%
\author{P. Saracco\inst{1}
\and M. Longhetti\inst{1}
\and P. Severgnini\inst{1}
\and R. Della Ceca\inst{1}
\and V. Braito\inst{1}
\and R. Bender\inst{2}
\and N. Drory\inst{3}
\and G. Feulner\inst{2}
\and U. Hopp\inst{2}
\and F. Mannucci\inst{4}
\and C. Maraston\inst{5}
}
\authorrunning{Paolo Saracco et al.}
%
%
\institute{INAF - Osservatorio Astronomico di Brera, Via Brera 28, 20121 Milano, Italy
\and Universit\"ats-Sternwarte M\"unchen, Scheiner Str. 1, 81679 M\"unchen, Germany
\and University of Texas at Austin, Austin, Texas 78712
\and IRA-CNR, Largo E. Fermi 5, 50125 Firenze, Italy
\and Max-Plank-Institut fuer  extraterrestrische Physik, Garching bei Munchen, Germany
}

\maketitle              


\section{Probing the formation of early-type galaxies}
How and when present-day massive early-type galaxies built up and
what type of evolution has characterized their growth
(star formation and/or merging)  still remain open issues.
The different competing scenarios of galaxy formation predict much 
different properties of early-type galaxies at $z>1$.
The ``monolithic'' collapse predicts that massive spheroids formed at
high redshift ($z>2.5-3$) and that their comoving density is constant 
at $z<2.5-3$ since they  evolve only in luminosity.
On the contrary, in the hierarchical scenario massive spheroids are built up 
through  subsequent mergers reaching their final masses at $z<1.5$
(\cite{kau1}, ~\cite{mou1}).
As a consequence,  massive systems are very rare at $z>1$, their
comoving density decreases from $z=0$ to $z\sim1.5$ and they
should experience their last burst of star formation at $z<1.5$, 
concurrent with the merging event(s) of their formation.
These opposed predicted properties of early-types at $z>1$ 
can be probed observationally once a well defined sample of massive 
early-types at $z>1$ is available.
We are constructing such a sample through a dedicated
near-IR very low resolution ($\lambda/\Delta\lambda\simeq50$)
spectroscopic survey (TNG EROs Spectroscopic Identification Survey, 
TESIS, ~\cite{sar1}) of a complete sample of 30 bright (K$<$18.5) 
Extremely Red Objects (EROs).

The sample has been selected over two fields ($\sim$360 arcmin$^2$) of
 the Munich Near-IR Cluster Survey 
(MUNICS,~\cite{mun1}) covered by B, V, R, I, J and 
K-band observations.
The red optical-to-near-IR colors (R-K$\ge$5.3) allow to select $z>1$ evolved 
stellar systems; the bright K magnitudes assure the selection of 
massive galaxies and the near-IR spectra allow the detection of the 4000Å 
break at $z>1$.
The survey is carried out at the 3.6m Italian Telescopio Nazionale Galileo 
(TNG) with the prism disperser Amici designed to cover the full 
spectral range (8000-25000 \AA) in a single shot, thus being very 
efficient in detecting continuum breaks and in describing spectral shapes.

In parallel with the near-IR spectroscopic follow-up we obtained 150 ks of 
XMM-Newton observations for the two selected fields (75 ks each) to study the 
nature of the X-ray emitting EROs. 
The results relevant to the first set of XMM observations 
are presented by Severgnini et al. (see contribution in this volume).

\section{The density of $\mathcal{M}_{star}$$>$$10^{11}M_\odot$ early-type
galaxies at $z>1$}
We  classified 10 out of the 13 EROs (40\%  of the sample) observed 
so far: 7 early-type galaxies 
and 3 starbursts.
The 7 early-type galaxies have apparent magnitudes 
$16.7\le K \le18.3$ and are at redshift $1.2<z<1.45$.
Their rest-frame K-band absolute magnitudes are in the range 
$-26.9\le M_K \le-25.7$, i.e. their  luminosities are $L>5L^*$
(we considered M$^*_K$=-24.2 from ~\cite{cole},  
H$_0$=70 Km s$^{-1}$ Mpc$^{-1}$, $\Omega_m=0.3$ and $\Omega_\lambda=0.7$).
In the hypothesis of passive evolution from $z\sim1.5$ to $z=0$
($\Delta K \simeq1$ mag), their resultant luminosities would be 
L$_{z=0}>$2L$^*$.
Thus, these  early-types are the $z>1$ counterpart of the  ellipticals 
populating the  bright end of the local luminosity function. 
Assuming a mass-to-light ratio $\mathcal{M}/$L$_K$=0.5 [M/L]$_\odot$, 
we derived a stellar mass well in excess to $2\times10^{11}M_\odot$
in all of them.
\begin{figure}
\vskip -0.5truecm
\begin{center}
\includegraphics[width=0.8\textwidth]{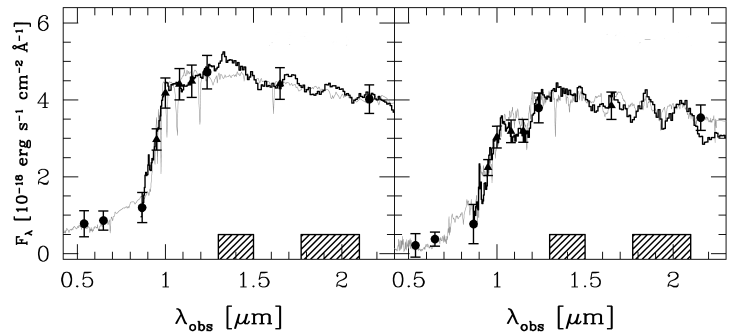}
\end{center}
\vskip -0.6truecm
\caption[]{Smoothed Amici spectra (black histogram) of two out of the 7 
early-type galaxies superimposed on synthetic spectra ($\tau<1$ Gyr and 
A$_V<0.5$ mag; thick gray line).
The circles are the photometric data in the V, R, I, J and K' bands
and the filled triangles are additional photometric points derived from 
the observed spectra (see ~\cite{sar1}).
Atmospheric windows with opacity larger than 80\% are marked by 
the shaded areas.
}
\vskip -0.5truecm
\end{figure}
These 7  early-types account for a comoving density of about 
2.6$\times10^{-5}$ Mpc$^{-3}$.
This is a lower limit to the density of 
$\mathcal{M}_{star}$$>$$10^{11}M_\odot$ early-types at $z>1$
since only 40\% of the sample of EROs has been observed.
Since, the number density of local L$_{z=0}>$2L$^*$ is 7$\times10^{-5}$ 
Mpc$^{-3}$ (we used $\phi^*_{E/S0}=1.5\times10^{-3}$ Mpc$^{-3}$, 
from ~\cite{marz}),  the comoving density of  massive ellipticals 
cannot vary more than a factor $\sim2$ from $z\sim1.5$ to $z=0$.
The completion of the survey will allow us to further refine
this value thus constraining severely  the evolution of 
massive ellipticals.

\end{document}